\documentstyle[prb,preprint,aps]{revtex}
\begin{document}
\draft
\title
{\bf Quantum heat transfer through an atomic wire }
\author{A.~Buldum$^{\ast}$ and S.~Ciraci} 
\address{
Department of Physics, Bilkent University, Bilkent 06533,
Ankara, Turkey.}
\author{C. Y.~Fong}
\address{
Department of Physics, University of California, Davis,
California 95616-8677, U.S.A}
\maketitle
\begin{abstract}
We studied the phononic heat transfer through an atomic 
dielectric wire
with both infinite and finite lengths by using a model
Hamiltonian approach. At low temperature 
under ballistic transport, the thermal conductance 
contributed by each phonon 
branch of a uniform and harmonic chain 
cannot exceed the well-known value which depends linearly on 
temperature but is material independent.
We predict that this ballistic thermal conductance 
will exhibit stepwise behavior as a function of temperature.
By performing numerical calculations on a more realistic 
system, where a  small atomic chain is placed between
two reservoirs, we also found resonance modes, 
which should also lead to the stepwise behavior in the thermal conductance. 
\end{abstract}
\narrowtext
\newpage
\section{ Introduction }
Many studies have demonstrated that transport properties of condensed 
matter with nanometer size range and low dimensionality can 
exhibit interesting features. For example, the electron transport
through a nanoobject (i.e. a small molecule, nanowire 
or atomic chain) shows novel features\cite{1,2,3,4,5,6,7,8} owing to 
two facts: (1) The electronic states confined
in directions which are perpendicular to the one for the current flow 
are quantized. The level spacings\cite{9}, 
$\Delta\epsilon_{i}=\epsilon_{i+1}-\epsilon_{i}$, of the quantized states 
of the nanoparticle are 
finite but strongly size and geometry dependent. (2) The mean distance, $l$, 
traveled by an electron between two terminals is smaller 
than the phase breaking length,  $L_{\phi}$, (i.e. $l < L_{\phi}$). 
Consequently, a stepwise behaviour in the electrical conductance
is observed under ballistic transport through a point contact 
created by an STM tip\cite{4,10}
(or through a narrow constriction between two reservoirs of
two-dimensional electron gas in high mobility
GaAs-GaAlAs heterostructures\cite{5,6}),
because the quantization and coherence of 
electronic states become pronounced. 
The ballistic electrical conductance $G$ was found\cite{5,6}  to vary with 
the radius of contact, $r_{c}$, (or with the width of the constriction,
$w$) in quantized steps of approximately $2e^{2}/h$, when $r_{c}$ or
$w$ is in the range of Fermi wave length $\lambda_{F}$. \\
\\
In view of the interesting physics underlying the ballistic electron 
transport through a nanoobject between two electrodes,  
we conjectured\cite{9} earlier that similar effects should occur also 
in the heat conduction {\it via phonons}.
Similar to a constriction in electrical conduction, we expect that
a nanoobject has a discrete vibrational frequency spectrum 
with finite spacings,
$\Delta \Omega_{i}=\Omega_{i+1}-\Omega_{i}$, determined by its atomic
configuration and size. Furthermore, when a 
nanoobject is placed between two
reservoirs, collective vibrational modes (phonons) which involve the
coherent vibrations of the atoms in the nanoobject and reservoirs
can be developed, leading to the process of ballistic thermal 
conductivity. The thermal conductance, $\cal K$,   
depends on the available number of channels and hence it is
expected to increase in discrete values each time a new ballistic
channel is included in the heat transfer. 
Each vibrational mode 
contributes to the heat transfer, but
its contribution is weighted
by Planck's distribution $n(\Omega, T)$. Therefore,
${\cal K}$ is determined by phonons with energies up to thermal energy 
($\hbar \Omega \sim k_{B}T$). This may lead to smearing of the
step structure of ${\cal K}$.
This feature of quantization of thermal conduction 
differs from that of electrical conduction,
because the former complies with 
Bose-Einstein statistics.\\            
\\
Heat transfer through a nanoobject is also related to
the atomic scale dry friction and lubrication\cite{11,12,13}. 
When two objects move relative to each other, they engage in contact through 
various atomic
scale asperities or lubricant molecules. 
In these studies, the emphasis is placed on the energy dissipation. 
The mechanical energy of 
the relative motion is 
dissipated by (a) the creation of electron-hole pairs, and (b) by
the creation of non-equilibrium phonon distribution at the close proximity 
of the asperity\cite{12,13}. 
In the latter case, the main issue is the thermal conductance related to
the heat transfer between the vibrational modes of the asperities
(or lubricant molecule) and those of the objects, that leads to
the energy dissipation\cite{14}.
In addition, the heat
transfer is also an interesting issue in nanodevices\cite{15}. 
Therefore, it is important to understand the thermal conduction through a 
nanoobject. \\ 
\\
The objective of this work is to investigate the effects of the quantized
nature of vibrational modes of a dielectric wire 
on {\it phononic} heat transfer between two reservoirs.
To this end, the ballistic heat transfer through
a uniform dielectric wire is investigated first. 
We also studied the thermal conduction through a finite atomic wire.
As an example, the heat transfer through a three-atom chain
placed between two reservoirs with large cross section are
examined by performing numerical calculations on a model system\cite{16}.
We found that
the thermal conductance of a uniform atomic wire 
at low temperature is independent
of material and increases linearly with temperature. For finite atomic wire
both longitudinal acoustic and optical modes contribute to the thermal
conductance; the latter modes may lead to stepwise behavior. 
\section{Model Hamiltonian Approach}
The system of interest is schematically described 
in the upper panel of Fig. 1. Heat flows from
the left (L) to the right (R) through a nanoobject (C).
Here $L$, $R$ can be sections of nanoobject itself or heat reservoirs.
The temperature of L and R, $T_{L}$ and $T_{R}$, respectively are
kept constant, so that the steady state of the transport is considered.
For this system we investigate the thermal conduction by using a model
Hamiltonian approach. To find the modes which are responsible for
the ballistic transport of heat, we specify first
a zero order vibrational Hamiltonian which characterizes the possible
modes in the three regions,$(L+C+R)$, 
\begin{equation}
H_{o}=\sum_{i,q_{L}} \hbar \Omega_{iq_{L}}a^{+}_{iq_{L}}a_{iq_{L}}
+ \sum_{i,q_{R}} \hbar \Omega_{iq_{R}}a^{+}_{iq_{R}}a_{iq_{R}}
+ \sum_{i,q_{C}} \hbar \Omega_{iq_{C}}a^{+}_{iq_{C}}a_{iq_{C}}  
\end{equation}
Here $q_{X}$ is the momentum quantum numbers
for vibrational mode $\Omega_{iq_{X}}$ belonging to the branch $i$
in one of the three regions ($X=L,R,C$), and $a_{iq_{X}}$   
($a^{+}_{iq_{X}}$) is the
corresponding annihilation (creation) operator. The heat
transfer via phonons is realized when $L$,$C$, and $R$ are coupled and 
$\Delta T=T_{L}-T_{R} > 0$. The coupling can be described by an 
interaction Hamiltonian,  
\begin{eqnarray}
H_{int}= \sum_{ij} {\cal T}_{ij,q_{L}q_{C}} (a^{+}_{jq_{C}}a_{iq_{L}}
+ a_{jq_{C}}a^{+}_{iq_{L}}) 
+ \sum_{jl} {\cal T}_{jl,q_{C}q_{R}} (a^{+}_{lq_{R}}a_{jq_{C}}
+ a_{lq_{R}}a^{+}_{jq_{C}}) \nonumber \\
+ \sum_{ijt} U_{ijt, q_{L}q_{C}q_{R}}( a_{iq_{L}}a^{+}_{jq_{C}}a^{+}_{tq_{R}}
+ \cdots
\end{eqnarray}
in terms of the harmonic $({\cal T})$ and anharmonic $(U)$ coupling 
coefficients. This Hamiltonian    
has been used to predict localized and extended  vibrational modes in
molecular crystals\cite{17}. The present application is similar to the
transfer Hamiltonian approach\cite{18}. As we emphasized earlier, we 
consider only the thermal conduction mediated by phonons. 
The electronic states, as
well as the electron-phonon interactions are not included 
in Eq.(1) and Eq.(2).\\
\\
The modes responsible for the ballistic heat transport are provided mainly
by the harmonic couplings. Before discussing the effects of the harmonic
coupling, we comment on the anharmonic terms.
The anharmonic couplings describe the phonon-phonon interactions 
which are the mechanism responsible for several dissipative
processes, such as energy loss to local modes of the nanowire and thermal
resistance due to umklapp process.  
Depending on the strength of the anharmonic coupling, 
the continuing energy transfer to the localized modes may result 
in {\it energy discharge} to the heat transporting states, 
or {\it desorption} or migration of an atom. 
The former event, i.e. the excitation of a local phonon,
followed by deexcitation can occur periodically and 
can be manifested in the time 
variation of the conductance, ${\cal K}(t)$. The detail of the anharmonic 
effects will not be discussed further in this paper. In what follows 
we apply the above approach to calculate the thermal conductance
of a uniform dielectric wire (Sec. III) and also of a finite atomic 
chain (Sec. IV). 
\section{Uniform dielectric wire}
We now consider three sections on a uniform wire.
Owing to the uniformity of the sections ($L,C,R$), the phonon modes
are the normal modes of the system, hence 
$q=q_{L}=q_{C}=q_{R}$, and also ${\cal T}_{ij,qq}={\cal T}_{ii,qq}$.  
Furthermore, ${\cal T}_{ii,qq}=\hbar\Omega_{iq}$. If the effect of
mode decay is considered, the rate of the heat transfer 
from $L$ to $R$ in this 1D system is given by 
$J^{LR}_{Q}=\sum_{iq} \hbar \Omega_{iq} \Gamma^{L}_{iq}$ 
by using the decay rate expression\cite{19} 
for a given $i$ and $q$,
\begin{equation}
\Gamma^{L}_{iq}=\Gamma^{C}_{iq}\frac{V^{LC}_{qq}V^{CL}_{qq}}
{(\hbar\Gamma^{C}_{iq})^2}
\end{equation}
with $\Gamma^{C}_{iq}=(2\pi/\hbar) V^{CR}_{qq}V^{RC}_{qq} 
D^{R}_{i}(\Omega_{iq})$.
By symmetry $V^{BA}_{qq}=V^{AB}_{qq}$ and
\begin{eqnarray}
V_{qq}^{LC}=\langle n_{L}-1, 
n_{C}+1, n_{R}|H_{int}|n_{L},n_{C},n_{R}\rangle  \nonumber \\
V_{qq}^{CR}=\langle n_{L}-1, n_{C}, 
n_{R}+1|H_{int}|n_{L}-1, n_{C}+1, n_{R}\rangle .
\end{eqnarray}
where $n_{X}$ ($X=L,C,R$) is the Planck's distribution
$n_{X}(\Omega_{iq},T_{X})=[exp(\hbar \Omega_{iq}/k_{B}T_{X})-1]^{-1}$
given in terms of the frequency of the mode and temperature of the section,
$T_{X}$.
Hence $V^{LC}_{qq}={\cal T}_{ii,qq}\sqrt{n_{L}(n_{C}+1)}$ and
$V^{CR}_{qq}={\cal T}_{ii,qq}\sqrt{(n_{C}+1)(n_{R}+1)}$  are expressed in 
terms of Planck's distribution functions. 
The net rate of heat transfer through 
the chain is $J_{Q}=J^{LR}_{Q}-J^{RL}_{Q}$.
Since $n_{L}$ and $n_{R}\ll 1$ at low temperature, $J_{Q}$ is 
expressed in the following form
\begin{equation}
J_{Q}\simeq\frac{\hbar}{2 \pi}\sum_{i}\int^{\infty}_{0}[ n_{L} - 
n_{R} ]\,\Omega \,d\Omega
\end{equation}
Here index $i$ of the summation sign specifies the branches for
$n_{L}$ and $n_{R}$. 
Note that Eq.(5) yields exactly the same expression of 
current one obtains from the
phenomenological approach\cite{20,21} by assuming the 
transmission coefficients, $t_{i}(\Omega)=1$.  
Therefore, the present approach
justifies the Landauer type expression of heat current by deriving it
from the model Hamiltonian specified by Eq.(1).
Now, we consider first an atomic
chain that has only three branches and 
evaluate Eq.(5) for $\Delta T \rightarrow 0$.
We take $T_{L}$ and $T_{R}$ small so that the upper limit of the integral
in Eq.(5)
is justified.
Defining $T=(T_{L}-T_{R})/2$, 
the following expression for the ballistic
thermal conductance, ${\cal K}=J_{Q}/\Delta T$, is obtained, 
\begin{equation}
{\cal K}=\sum_{i}\frac{\pi^{2} k^{2}_{B}}{3h}T
\end{equation}
According to this interesting result the ballistic conductance of each 
branch $i$ of the uniform and harmonic atomic chain is limited by the value 
${\cal K}_{o}=\pi^{2} k^{2}_{B}T/3h$. 
It is independent of any material property,
and is linearly dependent on $T$. 
The total thermal conductance becomes 
${\cal K}={\cal N}{\cal K}_{o}$, where ${\cal N}$ 
is the total number of branches. 
For an ideal 1D atomic chain ${\cal N}=3$, 
if the transverse vibrations are allowed.
The result agrees with the well-known expression 
derived earlier for dielectric
bridges\cite{22,23,24}. For a homogeneous wire that has a larger cross section,
more subbands can now be present due to the transverse confinement of the
vibrational motions in the direction perpendicular to the propagation.  
In this case, the ballistic thermal conductance of a branch
with $\Omega_{i,min}\neq 0$ is smaller than 
${\cal K}_{o}$ due to the finite frequency of a subband through the occupancy
of Planck's distribution. 
For example, with two subband frequencies,  
$\Omega_{min}=10^{10} Hz$ and $\Omega_{min}=10^{11} Hz$
at ${\bf q}=0$, the thermal conductance of each channel 
at $T=0.25K$ is calculated from Eq.(5) to be 
$0.94{\cal K}_{o}$ and $0.25{\cal K}_{o}$, respectively. 
At $T=1 K$, these contributions increase to $0.97 {\cal K}_{o}$ 
and $0.77 {\cal K}_{o}$, respectively. Consequently,   
one expects that each subband in the wire having
$\Omega_{i,min}\neq 0$, that formed by increasing the cross section,  
leads to a jump in ${\cal K}$ as in the electrical conduction. 
However, the amplitude of steps decreases 
exponentially with increasing $\Omega_{i,min}$ at given low temperature T. 
Furthermore, an important difference between the electrical and thermal 
conductance is 
that all the modes of a given branch $i$ 
contribute to ${\cal K}_{o}$, whereas the quantum ballistic electrical 
conduction in the 1D channels\cite{8,25} 
is governed by the Fermi-Dirac distribution. \\ 
\\
Recently, Greiner 
et al.\cite{26} calculated the thermal conductance of free
electrons for 1D ballistic and diffusive transport
in the linear response regime by using the correlation 
function formalism. They derived a universal expression
for the electronic thermal conductance of a degenerate 
electron gas, ${\cal K}^{el}_{o}$, that is the same as ${\cal K}_{o}$
found for the phononic thermal conductance in Eq.(6). Interestingly, 
electrons and phonons have the same universal
value for their thermal conductance 
at low temperature, in spite of the different statistics
they obey. We believe that 
this situation originates from the fact that both systems follow the same 
distribution,  
$e^{- \epsilon / k_{B} T}$ for $\epsilon /k_{B}T >>1$. Here, 
$\epsilon = \hbar \Omega_{i}$ for phonons, and 
$\epsilon = \epsilon_{i} - \mu$ for electrons, $\epsilon_{i}$ and $\mu$ being
electron energy and chemical potential.
\section{Finite wire and numerical modeling}
The quantum heat transfer through a uniform and harmonic atomic 
wire is an idealized situation. A more realistic system consists  
of a finite atomic wire, $C$ coupled 
to reservoirs $L$ and $R$. Here the finite atomic wire has discrete 
vibrational frequency
spectrum, while those of reservoirs are continuous. In the lower 
panel of Fig. 1 the corresponding densities of states, $D^{X}(\Omega)$, 
are schematically described. It is also noted that cross sections 
change abruptly where the atomic wire is connected to the reservoirs. 
As a result, the transmission coefficient,  $t_{j}(\Omega ) < 1$ because
of reflections. On the other hand, the application of 
the continuum theory to determine the transmission coefficients
as in the case of dielectric 
wires\cite{23} having cross section (50nm x 50nm) is not justified here 
owing to the discrete character and small size of the finite
atomic wire. 
Moreover, the modes of wire differ from those of the
reservoirs; $C$ itself is not uniform in the sense that the atoms
near the boundary experience different force constants. We now   
treat such a general non-uniform dielectric wire.  \\ 
\\
In order to analyze the character of the modes of
a finite wire and to present an example 
we carried out numerical calculations on a prototype 
model that consists of a chain of 3  
atoms placed between two reservoirs. Each 
reservoir ($L$ and $R$) is modeled by 7 (001) atomic planes of the
fcc structure. 
The whole system ($L+C+R$) 
comprises 451 atoms. The interatomic distances 
between an atom in the chain and nearest slab-atoms and 
between its adjacent chain atoms are taken, 2.7, 2.6\AA , respectively. 
We assume that the electrical conduction through these atoms is zero,
and hence the energy transfer through electrons is not allowed. 
Since we are concerned with the vibrational modes 
of the system, which are responsible for the energy transfer, 
we express the interatomic interactions by a prototype empirical,
Morse-type two-body potential. 
Then, we set up the dynamical matrix of the system and
calculate the normal modes and their frequencies $\Omega _{iq}$.\\ 
\\
Our analysis distinguished three different types of modes: i) Modes localized
strongly in the chain. ii) Modes confined to the reservoir(s).
iii) Current-transporting resonant modes.
Note that only confined modes would appear if the coupling
between the chain and the reservoirs were weak. The modes
in the first two categories ( i, ii) cannot be involved in the
heat transfer, but they affect the thermal conduction 
by increasing the thermal resistance through anharmonic couplings.
The transverse modes of the finite chain are well localized owing
to the different coupling constants and different 
number of neighbors as compared to the atoms in
the reservoirs. On the other hand, the longitudinal modes are
coupled to those of the reservoirs and form resonant 
modes. Figure 2 illustrates three resonant modes of the atoms. 
Their relative motions are indicated by the arrows. (a) and (b)
can be considered as longitudinal optical branches and (c) the acoustic 
branch. Therefore, the resonant modes can be also contributed
by the optical branches. Figure 2 also shows that 
the local densities of states within the chain,
$\rho_{iq_{C}}(\Omega )$, associated with these modes,
$iq_{C}$, which are calculated from the spectrum of the dynamical matrix.
We now consider the ballistic heat transfer through 
resonances. Starting from the rate expression given in Sec. III, 
the ballistic thermal conductance for 
a single resonant mode $iq$, is given by,
\begin{equation}
\kappa_{iq}\sim k_{B}v_{iq}x^{2}e^{x}(e^{x}-1)^{-2}
\end{equation}
with $x=\hbar\Omega_{iq}/k_{B}T$. For large $x$, $\kappa_{iq}\sim
v_{iq}k_{B}x^{2}e^{-x}$. On the other hand 
$\kappa_{iq}\sim v_{iq}k_{B}$ if $x$ is small. 
In view of the analysis in Ref [25], the thermal conductance 
of a finite atomic chain 
due to all resonance modes can be obtained from the sum over these  
modes, $iq_{C}$, 
\begin{equation}
{\cal K}\sim \sum_{iq_{C}} v_{iq_{C}} k_{B} \int_{0}^{\infty}
\rho_{iq_{C}}(\Omega ) \left( \frac{\hbar\Omega }
{k_{B}T} \right)^{2}exp\left(\frac{\hbar\Omega }{k_{B}T}\right)
\left[exp\left(\frac{\hbar\Omega }{k_{B}T}\right)-1\right]^{-2}\,d\Omega 
\end{equation}
Here $\rho_{iq_{C}}$ is the local density of states for the resonant 
modes derived from the mode $iq_{C}$ and $v_{iq_{C}}$ is average velocity
associated with the same mode. This density of states modifies
the ideal ballistic transmission as $t_{iq}(\Omega)$ does to $J_{Q}$
in Ref [23]. If $\rho_{iq_{C}}$ in Eq.(8) is replaced by 
$D_{i}^{C}(\Omega )$ corresponding to a branch of uniform atomic chain,  
the thermal conductance 
${\cal K}$ in Eq.(8) yields the universal value ${\cal K}_{o}$. 
By using a value of $v_{iq_{C}}\sim 393 m/sec$ deduced from the
slab calculations, we calculate 
${\cal K}(T)$ due to resonance in Fig. 2(c) and illustrate it
in inset in the same panel. For $T<< 50$ K, ${\cal K}$ increases with $T$,
then ${\cal K}$ saturates as $T \rightarrow 50$ K. This latter fact
is the consequence of $\kappa_{iq}$ at small $x$ discussed above.   
\\
We shall emphasize an interesting feature of 
${\cal K}$ that can be revealed from Eq.(8). Each 
atom incorporated in the finite chain between two reservoirs
adds 3 (optical) modes. So as the chain becomes longer 
by including new atoms, the number of longitudinal optical modes 
which may give rise to resonant modes increases. Accordingly, each 
resonant mode labeled by $iq_{C}$ that can lead to 
a current transporting channel,
shall give rise to a step increase in ${\cal K}$ by increasing the temperature. 
Apparently, the step behavior can occur without having subbands in the
transverse direction. This situation is relevant for the energy 
transport through molecules attached (i.e chemisorbed or physisorbed)
to the surfaces. The thermal conductance associated with the phononic
heat transfer for a molecule having discrete vibrational frequency
spectrum may exhibit at least non-monotonic variation with temperature.
Whether the step behavior of thermal conductance can be
resolved depends on the 
frequency and level spacing of the resonance modes. 
According to a criterion put forward by Blencowe\cite{24},
$\Omega_{i+1} > 14 \Omega_{i}$ in order to resolve the quantum features
of ${\cal K}(T)$.
As for the question whether ${\cal K}$ of a molecule or an atomic
chain between two reservoirs increases by incorporating new atoms,
the answer is not straightforward. It depends on the new modes yielding
new resonance mode, and also $\rho_{iq_{C}}(\Omega)$ and $v_{iq_{C}}$
associated these modes. Nevertheless, this is an interesting
issue that deserves further study.  
The contacts to the reservoirs by themselves can set a resistance.
It should also be noted that increasing the length  
of the chain by including more atoms can lead to 
the possibility of introducing imperfections in the structure.
The phonon-phonon interaction due to anharmonic coupling 
(that is not considered here) can also contribute to the resistance 
of the wire.
Hence the probability of observing such coherent conductance decreases.
As a result, the resistance to heat current increases.
\section{Conclusions}
In this work we analyzed important features of the heat 
conduction via phonons through a dielectric wire. 
In particular, we investigated 
the ballistic heat transfer through 
a uniform and harmonic wire and also through a finite 
atomic chain. By using a model Hamiltonian approach we showed that 
the conductance of a uniform atomic chain at low temperature 
is consistent with the well-known universal value, and varies linearly  
with temperature. As the cross section of the wire becomes larger to 
include several subbands due to the confinement of the transverse 
motion of ions, each subband contributes to the ballistic thermal 
conductance by an amount 
that depends on $T$ and the threshold frequency 
of the subband, $\Omega_{i,min}$. 
The higher the frequency of the mode, the less 
its contribution to ${\cal K}$. The resulting step behavior shall be 
observable, if they are not smeared out. We analyzed also 
the nature of modes for a finite atomic chain placed between two reservoirs. 
We found resonance modes originating from longitudinal optical modes, 
which can contribute to the ballistic thermal conductance. 
They differ from the subbands due to lateral confinement but should also
exhibit stepwise behavior in ${\cal K}$. 

\acknowledgments

We acknowledge stimulating discussions with Dr. D. M. Leitner. We
thank Mr Sefa Dag and A. Ozpineci  
for their assistance in the preparation
of the manuscript. This work is supported in part by the grants
NSF-INT-9872053 and TUBITAK.

$^{\ast}$ Present Address: Department of Physics and Astronomy, 
The University of North Carolina 
at Chapel Hill, Chapel Hill, NC, 27599

\begin{figure}
\caption{
A schematic description of the quantum heat transfer investigated in
the text. A nanoobject coupled to two reservoirs and corresponding 
density of frequencies. Resonance states are described by dotted lines.}
\end{figure}

\begin{figure}
\caption{
The local density states corresponding to resonance 
modes derived from the three 
longitudinal modes $iq_{C}$'s of a chain of three atoms between 
two slabs representing the reservoirs are illustrated in panels
(a)-(c). The corresponding modes are schematically
described at right. In (c) the 
variation of ${\cal K}$ 
in units of $10^{-21} Jm/Ksec$ calculated from the acoustic resonance modes 
is illustrated in the same panel.} 
\end{figure}

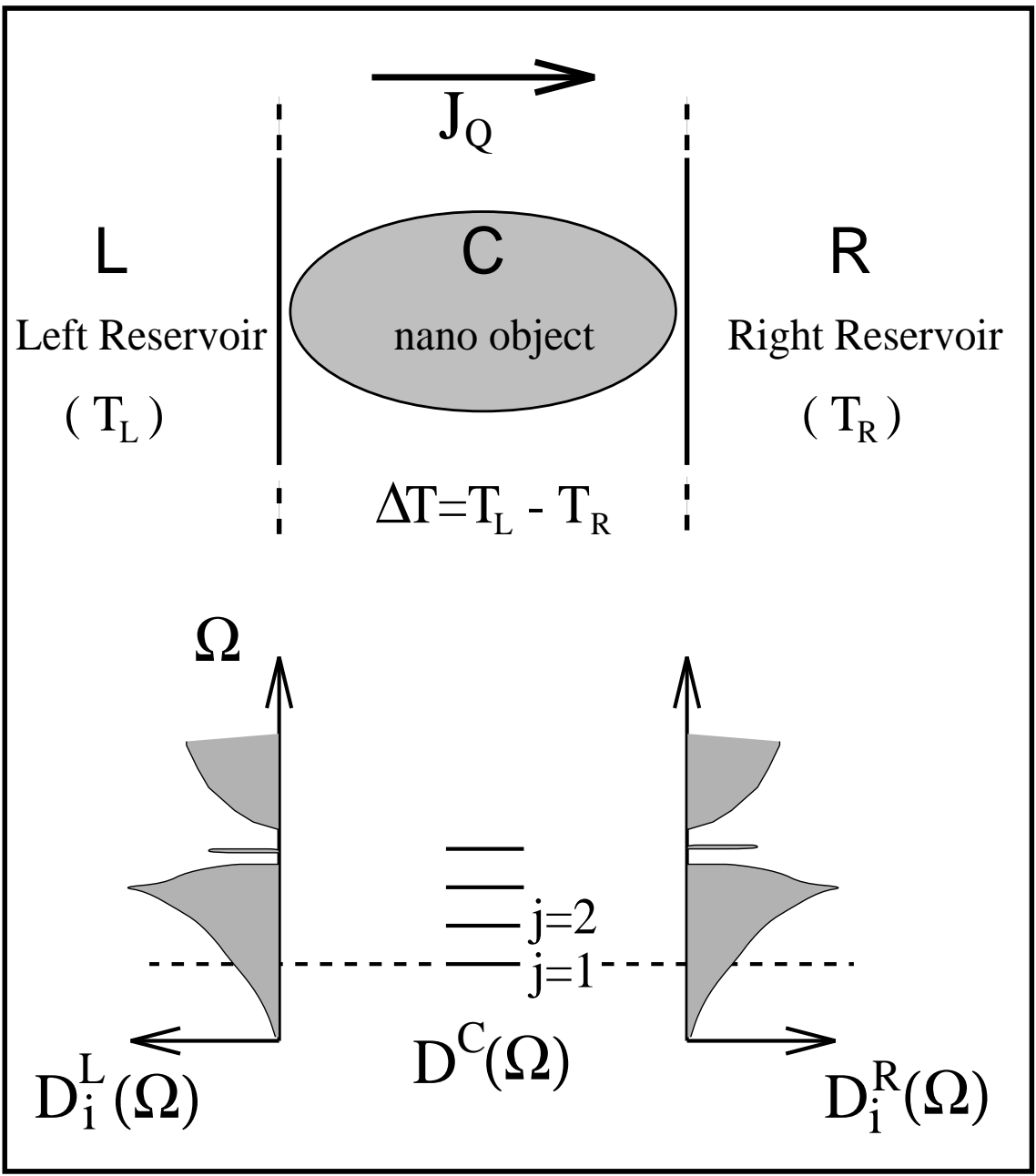
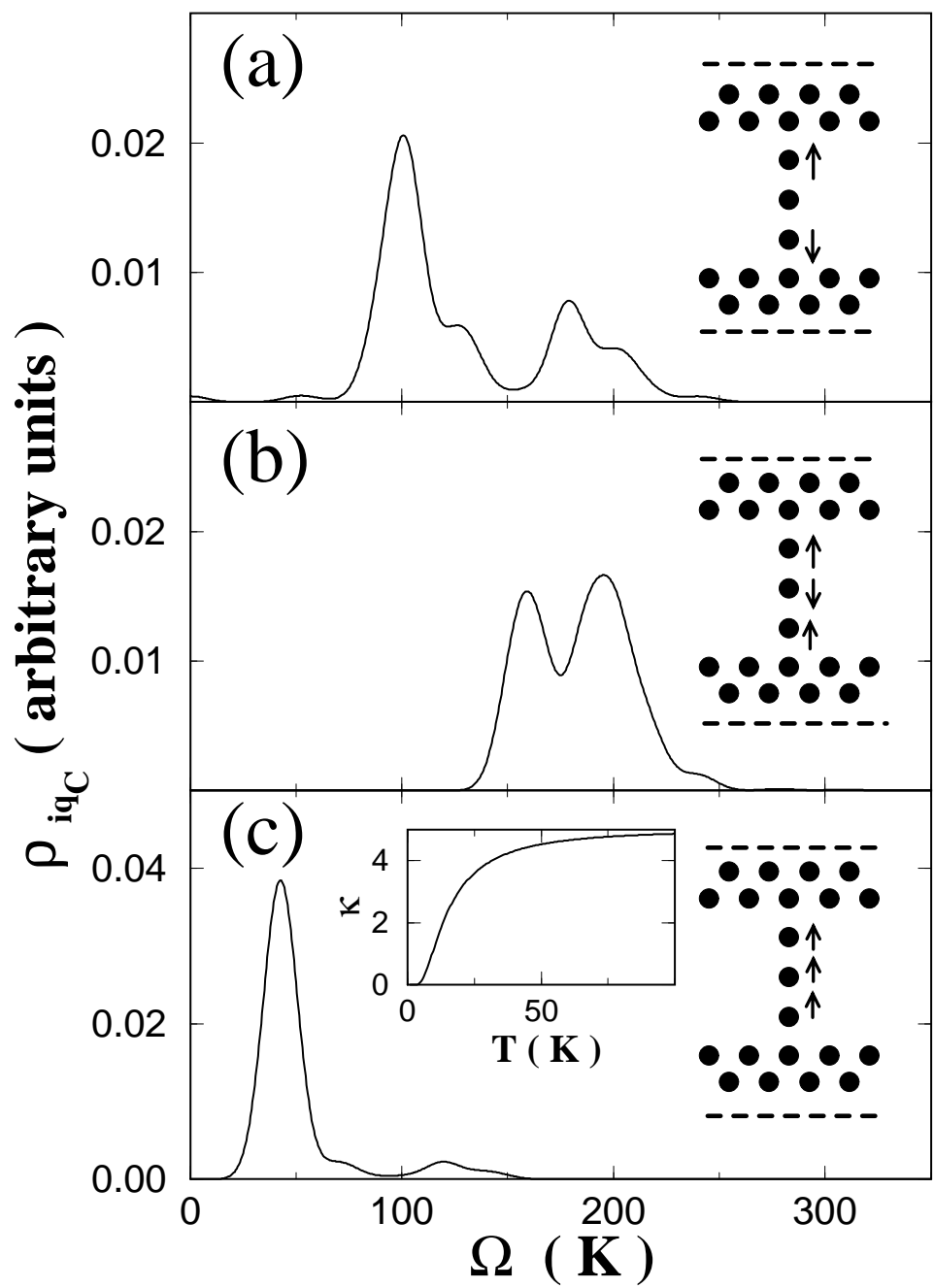
\end{document}